\documentclass[aps,twocolumn,prb,showpacs,amsmath,amssymb,superscriptaddress,letterpaper]{revtex4}
\usepackage{times}
\usepackage{amsfonts}
\usepackage{mathrsfs}
\usepackage{graphicx}% Include figure files
\usepackage{dcolumn}% Align table columns on decimal point
\usepackage{bm}% bold math
\usepackage{color}

\usepackage[colorlinks,bookmarks=false,citecolor=blue,linkcolor=red,urlcolor=blue]{hyperref}
\def\be{\begin{equation}}       \def\ee{\end{equation}}
\def\bea{\begin{eqnarray}}      \def\eea{\end{eqnarray}}

\begin{document}
\title{Topology and symmetry of circular photogalvanic effect in the chiral multifold semimetals: a review}

\author{Congcong Le}\email{Corresponding: Congcong.Le@cpfs.mpg.de}
\affiliation{Max Planck Institute for Chemical Physics of Solids, 01187 Dresden, Germany}
\affiliation{RIKEN Interdisciplinary Theoretical and Mathematical Sciences (iTHEMS), Wako, Saitama 351-0198, Japan}
%
%\author{Claudia Felser}
%\affiliation{Max Planck Institute for Chemical Physics of Solids, 01187 Dresden, Germany}
%

\author{Yan Sun}\email{Corresponding: ysun@cpfs.mpg.de}
\affiliation{Max Planck Institute for Chemical Physics of Solids, 01187 Dresden, Germany}

\date{\today}

\begin{abstract}
The circular photogalvanic effect (CPGE) is the only possible quantized signal in Chiral Weyl and
 multifold semimetals with inversion and mirror symmetries broken. Here, we review CPGE in the chiral multifold semimetals in terms of classification of CPGE tensor, the quantization of CPGE from $k\cdot p$ effective  model and topological semimetal RhSi family. Firstly, we give complete symmetric analysis of CPGE tensors for all nonmagnetic point groups, and get a table classifying matrix of response tensors. Secondly, the CPGE becomes a quantized response in the noncentrosymmetric topological semimetals, and depends on the Chern number of multifold fermions. Based on $k \cdot p$ effective model with linear dispersion, detailed derivations about the quantization of CPGE are given. Finally, according to ab-initio analysis for the quantized CPGE based on noninteracting electronic structures, we review previous reports and make new calculations for the chiral topological semimetals in RhSi family, which can be separated into two groups. The first group, including RhSi, PtAl and CoSi, can be the promising candidates to exhibit a quantized CPGE trace, while the second group includes PdGa, PtGa and RhSn without a quantization.
\end{abstract}


\maketitle

\section{Introduction}

The nonlinear optical responses, such as photogalvanic effects, play a crucial role not only in optical devices but also in probing the fundamental properties of quantum crystal materials\cite{Bloembergen1996,Boyd2008,Belinicher1980,Sturman1992}. According to the polarization state of the photons that induce the photocurrents, the photogalvanic effects can be classified into linear photogalvanic effect (LPGE) and circular photogalvanic effect (CPGE). The LPGE is observed with linearly polarized light, while the CPGE can be generated by circularly polarized light. A characteristic feature of CPGE is that the photocurrent can reverse its direction upon changing the radiation helicity from left-handed to right-handed\cite{Belinicher1980}. In general, the CPGE photocurrent can be realized in several different systems: (1) The first one is conventional bulk material without inversion symmetry\cite{Ivchenko1978,Belinicher1978,Asnin1978}. (2) The second one is quantum wells with strong Rashba spin-orbital couples (SOC), where the spin structure of Rashba SOC can lead to CPGE photocurrents under circularly polarized radiation\cite{Ganichev2003}. (3) The third one is the nontrivial surface states of topological insulators, and the CPGE photocurrent originates from  topological helical Dirac fermions\cite{McIver2011,Hosur2011}. (4) The fourth one is topological Weyl and unconventional fermions\cite{Ma2017,Juan2017,Chang2017,Felix2018}, where unconventional
fermions contain three-, four-, six- and eightfold degenerate points and these fermions can give rise to CPGE photocurrents. Interestingly, it has been pointed out \cite{Juan2017,Chang2017,Felix2018}
that a single multifold degenerate point with nontrivial topological invariant can lead to the quantization of CPGE, which can directly measure the topological charge of degenerate points.

In the topological semimetals\cite{Chiu2016,Armitage2018,Bradlyn2016,HgCrSe,XG Wang,intermediate phase,multilayer weyl,xu2015,Lv2015,Weng2015,Shekhar2015,Yang2015,Xu2015,Li29020,Dirac Kane,Na3Bi,Cr3As2,le2017,le2018}, the Nielsen-Ninomiya Theorem\cite{Nielsen1,Nielsen2,Nielsen3} has set up a ground rule for topological degenerate points in a Brillouin zone (BZ). The total topological charges in the entire BZ must be neutralized, and hence a topological degenerate point inevitably accompanies another topological point with opposite topological charge. The total CPGE contributed by all degenerate points is zero when degenerate points with opposite topological charges are the same energy due to crystal symmetries. In all crystal symmetries, only inversion and mirror symmetries can change the sign of topological charges, which can ensure that the degenerate points with opposite chirality have the same energy. The topological semimetals without inversion and mirror symmetries belong to chiral space group, only including time-reversal and rotational symmetries, and degenerate points with opposite topological topological charges can have different energy. In a certain frequency range, degenerate points with opposite charge do not contribute to CPGE at the same time, which can lead to the quantization of CPGE due to Pauli blocking. Therefore, the nonmagnetic chiral topological semimetals with chiral space group can host a quantized CPGE trace.

In the chiral topological semimetals, the RhSi family\cite{Chang2017,Tang2017,Rees2020,Ni2020,Shinoda1972,Narozhnyi2013,Geller1954,Takizawa1988,Larchev1982}, including CoSi, PdGa, PtGa, PtAl and RhSn, can be the promising candidates to exhibit a quantized CPGE. Firstly, the RhSi family belong to chiral space group G=P2$_1$3 (No.198) with  point group $T$, which only includes rotational symmetries. Secondly, the crystal symmetries and time-reversal symmetry can protect fourfold and sixfold fermions at $\Gamma$ and $R$ point, respectively. Thirdly, a big energy difference between fourfold and sixfold fermions has been confirmed by ARPES measurements\cite{Daniel2019,Rao2019}, and the fourfold degenerate point is near Fermi level. Hence, we will perform an ab initio analysis for the CPGE in RhSi family based on noninteracting electronic structures

The outline of this review is organized as follows. In Section.~\ref{S1},
the complete symmetry analysis of CPGE tensor are given for all nonmagnetic point groups. Then, in Section.~\ref{S2},
Based on $k \cdot p$ effective models with linear dispersion, detailed derivations of quantized CPGE trace  are discussed. In Section.~\ref{S3},
we review previous reports\cite{Le2020} and make some new calculations for the chiral topological semimetals in RhSi family, which can be separated into two groups. The first group can be the promising candidates to exhibit a nearly quantized CPGE trace, while the second group is far away from quantization. Finally, in Section.~\ref{S4}, we give a summary of our paper.

\begin{figure*}
\centerline{\includegraphics[width=0.9\textwidth]{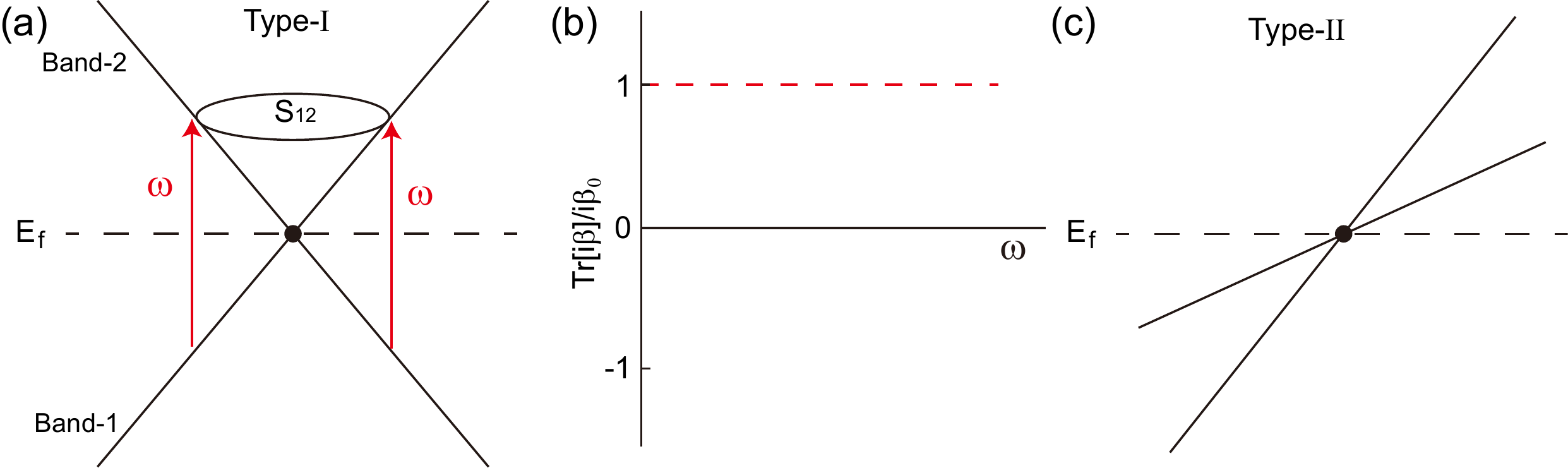}}
\caption{(color online) (a) Band structure of Type-I Weyl point in the two-band effective $k\cdot p$ model, where the surface $\boldsymbol{S}_{12}$ is closed and wraps Weyl point. (b) The trace of CPGE tensor as a function of frequency $\hbar\omega$ in the Type-I Weyl point. (c) Band structure of Type-II Weyl point. The chemical potential $E_f$ is indicated by by a dashed horizontal line.
\label{fig1} }
\end{figure*}

\section{complete Symmetry analysis of CPGE tensor}\label{S1}

In this section, based on the analysis method in the Ref.\cite{Le2020}, which only gives symmetry analysis of $T$ point group in the chiral multifold semimetal RhSi, here we will give more detailed symmetry analysis for all nonmagnetic point groups. The CPGE injection current and CPGE tensor $\boldsymbol{\beta}(\omega)$  can be written as\cite{Juan2017,Sipe2000}

\begin{eqnarray}
\frac{\mathrm{d} j_{i}}{\mathrm{d} t}=\beta_{i j}(\omega)\left[\boldsymbol{E}(\omega) \times \boldsymbol{E}^{*}(\omega)\right]_{j},
\beta_{i j}(\omega)=\sum_{\boldsymbol{k}}\tilde{\beta}_{i j}(\boldsymbol{k},\omega),\nonumber
\\
\Gamma_{ikl}(\boldsymbol{k},\omega)=\frac{\pi e^{3}}{\hbar^2 V}\sum_{ n, m} f_{n m}^{\boldsymbol{k}} \Delta_{\boldsymbol{k}, n m}^{i} r_{\boldsymbol{k}, n m}^{k} r_{\boldsymbol{k}, m n}^{l} \delta(\hbar \omega-E_{\boldsymbol{k}, m n}),
\label{eq1}
\end{eqnarray}

Where $\tilde{\beta}_{i j}(\boldsymbol{k},\omega)=\epsilon_{j k l}\Gamma_{ikl}(\boldsymbol{k},\omega)$, $\boldsymbol{E}^{*}(\omega)=\boldsymbol{E}(-\omega)$ is the electric field of circularly polarized light. $i$ and $j$ indexes are the direction of current and circular polarized light, respectively. $E_{\boldsymbol{k}, m n}=E_{\boldsymbol{k}, m}-E_{\boldsymbol{k}, n}$ and $f_{n m}^{\boldsymbol{k}}=f_{n}^{\boldsymbol{k}}-f_{m}^{\boldsymbol{k}}$ are differences between band energies and Fermi-Dirac distributions, $\Delta_{\boldsymbol{k}, n m}^{i}=\partial_{k_{i}} E(\boldsymbol{k})_{n m}$, and $r^{i}_{\boldsymbol{k}, n m}=i\left\langle m(\boldsymbol{k})\left|\partial_{k_{i}}\right| n(\boldsymbol{k})\right\rangle$. Under the crystal symmetry $g$, the relation of $\Delta_{g\boldsymbol{k}, n m}^{i}$ and $r^i_{g\boldsymbol{k}, n m}$ between $\boldsymbol{k}$ and $g\boldsymbol{k}$ can be written as

\begin{eqnarray}
\Delta_{g\boldsymbol{k}, n m}^{i}=\sum_{i^{\prime}}\frac{\partial (g\boldsymbol{k})_{i^{\prime}}}{\partial k_{i}}\Delta_{\boldsymbol{k}, n m}^{i^{\prime}},
r^i_{g\boldsymbol{k}, n m}=\sum_{i^{\prime}}\frac{\partial (g\boldsymbol{k})_{i^{\prime}}}{\partial k_{i}}r^{i^{\prime}}_{\boldsymbol{k}, n m},
\label{eq2}
\end{eqnarray}

Substitute Eq.\ref{eq2} into Eq.\ref{eq1}, we have the relation of $\tilde{\beta}_{i j}(\boldsymbol{k},\omega)$ and $\tilde{\beta}_{i j}(g\boldsymbol{k},\omega)$ connected by crystal symmetry $g$

\begin{eqnarray}
\tilde{\beta}_{i j}(g\boldsymbol{k},\omega)=\epsilon_{j k l}\sum_{i^{\prime},k^{\prime},l^{\prime}}\frac{\partial (g\boldsymbol{k})_{i^{\prime}}}{\partial k_{i}}\frac{\partial (g\boldsymbol{k})_{k^{\prime}}}{\partial k_{k}}\frac{\partial (g\boldsymbol{k})_{l^{\prime}}}{\partial k_{l}}\Gamma_{i^{\prime}k^{\prime}l^{\prime}}(\boldsymbol{k},\omega)
\end{eqnarray}

If we consider the systems with inversion symmetry $I$, and the tensor $\tilde{\beta}_{i j}(I\boldsymbol{k},\omega)$ becomes

\begin{eqnarray}
\tilde{\beta}_{ij}(I\boldsymbol{k},\omega)&=&\epsilon_{j k l}\sum_{i^{\prime},k^{\prime},l^{\prime}}\frac{\partial (I\boldsymbol{k})_{i^{\prime}}}{\partial k_{i}}\frac{\partial (I\boldsymbol{k})_{k^{\prime}}}{\partial k_{k}}\frac{\partial (I\boldsymbol{k})_{l^{\prime}}}{\partial k_{l}}\Gamma_{i^{\prime}k^{\prime}l^{\prime}}(\boldsymbol{k},\omega),\nonumber
\\
&=&\epsilon_{j k l}\frac{\partial (I\boldsymbol{k})_{i}}{\partial k_{i}}\frac{\partial (I\boldsymbol{k})_{k}}{\partial k_{k}}\frac{\partial (I\boldsymbol{k})_{l}}{\partial k_{l}}\Gamma_{ikl}(\boldsymbol{k},\omega),\nonumber
\\
&=&-\epsilon_{j k l}\Gamma_{ikl}(\boldsymbol{k},\omega)=-\tilde{\beta}_{ij}(\boldsymbol{k},\omega)
\end{eqnarray}

Where the second line uses $I\boldsymbol{k}=(-\boldsymbol{k}_x,-\boldsymbol{k}_y,-\boldsymbol{k}_z)$. Due to CPGE tensor $\beta_{i j}(\omega)=\sum_{\boldsymbol{k}}\tilde{\beta}_{i j}(\boldsymbol{k},\omega)$, we can get $\beta_{i j}(\omega)=0$. Therefore, the CPGE tensor $\beta_{i j}(\omega)$ in inversion symmetric systems is zero, and the materials with non-zero CPGE tensor should belong to the point groups without inversion symmetry. Table.\ref{beta} lists all point groups without inversion symmetry, which consist of rotational and mirror symmetries, and the principal axis is along the z direction. In the following, we will take point groups $C_{n}$(n=2,3,4,5) and $C_{1h}$ as examples to analyze CPGE tensor $\beta_{i j}(\omega)$, and then the other point groups can be obtained by these point groups.

Firstly, we consider point groups $C_{2}$ and $C_{1h}$, where the generators are $C_{2z}$ rotational symmetry and $M_{xy}$ mirror symmetry. Similar to inversion symmetry, we can get

\begin{eqnarray}
\tilde{\beta}_{ii}(C_{2z}\boldsymbol{k},\omega)=\tilde{\beta}_{ii}(\boldsymbol{k},\omega),\tilde{\beta}_{xy}(C_{2z}\boldsymbol{k},\omega)=\tilde{\beta}_{xy}(\boldsymbol{k},\omega),
,\nonumber
\\
\tilde{\beta}_{xz}(C_{2z}\boldsymbol{k},\omega)=-\tilde{\beta}_{xz}(\boldsymbol{k},\omega),
\tilde{\beta}_{yz}(C_{2z}\boldsymbol{k},\omega)=-\tilde{\beta}_{yz}(\boldsymbol{k},\omega),
\nonumber
\\
\tilde{\beta}_{ii}(M_{xy}\boldsymbol{k},\omega)=-\tilde{\beta}_{ii}(\boldsymbol{k},\omega),
\tilde{\beta}_{xy}(M_{xy}\boldsymbol{k},\omega)=-\tilde{\beta}_{xy}(\boldsymbol{k},\omega),
\nonumber
\\
\tilde{\beta}_{xz}(M_{xy}\boldsymbol{k},\omega)=\tilde{\beta}_{xz}(\boldsymbol{k},\omega),\tilde{\beta}_{yz}(M_{xy}\boldsymbol{k},\omega)=-\tilde{\beta}_{yz}(\boldsymbol{k},\omega)
\end{eqnarray}

Because of CPGE tensor $\beta_{i j}(\omega)=\sum_{\boldsymbol{k}}\tilde{\beta}_{i j}(\boldsymbol{k},\omega)$, the matrix of tensor $\beta_{i j}(\omega)$ can be written as

\begin{eqnarray}
C_{2}:\left(\begin{array}{ccc}
 \beta_{xx}(\omega) & \beta_{xy}(\omega) & 0\\
 \beta_{yx}(\omega) & \beta_{yy}(\omega) & 0\\
 0 & 0 & \beta_{zz}(\omega)
\end{array}\right),\nonumber
\\
C_{1h}:\left(\begin{array}{ccc}
0 & 0 &  \beta_{xz}(\omega)\\
0 & 0 &  \beta_{yz}(\omega)\\
\beta_{zx}(\omega) &  \beta_{zy}(\omega) & 0
\end{array}\right)
\end{eqnarray}

Secondly, we consider point groups $C_{4}$ and $C_{6}$, where the generators are $C_{4z}$ and $C_{6z}$ rotational symmetries. Since $C_{4}$ and $C_{6}$ include $C_{2z}$ rotational symmetry, and hence $\beta_{xz}(\omega)=\beta_{yz}(\omega)=\beta_{zx}(\omega)=\beta_{zy}(\omega)=0$, suggesting that these matrix terms can not be considered. Let's use $\tilde{\beta}_{xx}(C_{4z}\boldsymbol{k},\omega)$ as an example, which becomes

\begin{eqnarray}
&&\tilde{\beta}_{xx}(C_{4z}\boldsymbol{k},\omega)\nonumber
\\
&=&\epsilon_{x k l}\sum_{i^{\prime},k^{\prime},l^{\prime}}\frac{\partial (C_{4z}\boldsymbol{k})_{i^{\prime}}}{\partial k_{x}}\frac{\partial (C_{4z}\boldsymbol{k})_{k^{\prime}}}{\partial k_{k}}\frac{\partial (C_{4z}\boldsymbol{k})_{l^{\prime}}}{\partial k_{l}}\Gamma_{i^{\prime}k^{\prime}l^{\prime}}(\boldsymbol{k},\omega),\nonumber
\\
&=&\epsilon_{xyz}\frac{\partial (C_{4z}\boldsymbol{k})_{y}}{\partial k_{x}}\frac{\partial (C_{4z}\boldsymbol{k})_{x}}{\partial k_{y}}\frac{\partial (C_{4z}\boldsymbol{k})_{z}}{\partial k_{z}}\Gamma_{yxz}(\boldsymbol{k},\omega)
\nonumber
\\
&+&\epsilon_{xzy}\frac{\partial (C_{4z}\boldsymbol{k})_{y}}{\partial k_{x}}\frac{\partial (C_{4z}\boldsymbol{k})_{z}}{\partial k_{z}}\frac{\partial (C_{4z}\boldsymbol{k})_{x}}{\partial k_{y}}\Gamma_{yzx}(\boldsymbol{k},\omega),\nonumber
\\
&=&-\Gamma_{yxz}(\boldsymbol{k},\omega)+\Gamma_{yzx}(\boldsymbol{k},\omega)=\tilde{\beta}_{yy}(\boldsymbol{k},\omega)
\end{eqnarray}

Where  the second line is due to $C_{4z}\boldsymbol{k}=(-\boldsymbol{k}_y,\boldsymbol{k}_x,\boldsymbol{k}_z)$. The other matrix terms of tensor $\beta_{i j}(\omega)$ in the point groups $C_{4}$ and $C_{6}$ can be obtained as

\begin{eqnarray}
\tilde{\beta}_{xx}(C_{4z}\boldsymbol{k},\omega)=\tilde{\beta}_{yy}(\boldsymbol{k},\omega),~~\tilde{\beta}_{zz}(C_{4z}\boldsymbol{k},\omega)=\tilde{\beta}_{zz}(\boldsymbol{k},\omega),
\nonumber
\\
\tilde{\beta}_{xy}(C_{4z}\boldsymbol{k},\omega)=-\tilde{\beta}_{yx}(\boldsymbol{k},\omega),
\tilde{\beta}_{xx}(C_{6z}\boldsymbol{k},\omega)=\tilde{\beta}_{yy}(\boldsymbol{k},\omega),\nonumber
\\
\tilde{\beta}_{zz}(C_{6z}\boldsymbol{k},\omega)=\tilde{\beta}_{zz}(\boldsymbol{k},\omega),~~\tilde{\beta}_{xy}(C_{6z}\boldsymbol{k},\omega)=-\tilde{\beta}_{yx}(\boldsymbol{k},\omega),
\end{eqnarray}

The matrix of tensor $\beta_{i j}(\omega)$ in the point groups $C_{4}$ and $C_{6}$ are the same, which can be written as

\begin{eqnarray}
\left(\begin{array}{ccc}
 \beta_{xx}(\omega) & \beta_{xy}(\omega) & 0\\
 -\beta_{xy}(\omega) & \beta_{xx}(\omega) & 0\\
 0 & 0 & \beta_{zz}(\omega)
\end{array}\right)
\end{eqnarray}

Thirdly, we consider point group $C_{3}$ with generator of $C_{3z}$ rotational symmetry. Similar to above  derivations, under $C_{3z}$ rotational symmetry, we can get

\begin{eqnarray}
&\tilde{\beta}_{xx}(C_{3z}\boldsymbol{k},\omega)=\tilde{\beta}_{yy}(\boldsymbol{k},\omega),~~\tilde{\beta}_{zz}(C_{3z}\boldsymbol{k},\omega)=\tilde{\beta}_{zz}(\boldsymbol{k},\omega),
\nonumber
\\
&\tilde{\beta}_{xy}(C_{3z}\boldsymbol{k},\omega)=-\tilde{\beta}_{yx}(\boldsymbol{k},\omega),
\end{eqnarray}

The corresponding matrix of tensor $\beta_{i j}(\omega)$ is

\begin{eqnarray}
\left(\begin{array}{ccc}
 \beta_{xx}(\omega) & \beta_{xy}(\omega) & 0\\
 -\beta_{xy}(\omega) & \beta_{xx}(\omega) & 0\\
 0 & 0 & \beta_{zz}(\omega)
\end{array}\right)
\end{eqnarray}

Finally, based on the CPGE tensor $\beta_{i j}(\omega)$ of above point groups, other point groups can be obtained. For examples, the point group $C_{2v}$ can be written as $C_2\otimes C^{\prime}_{1h}$,  where the generators are $C_{2z}$ rotational symmetry and $M_{xz}$ mirror symmetry. Similar to $C_{1h}$ with $M_{xy}$ mirror symmetry, tensor matrix of $C^{\prime}_{1h}$ is

\begin{eqnarray}
 C^{\prime}_{1h}:\left(\begin{array}{ccc}
0 & \beta_{xy}(\omega) &  0\\
\beta_{yx}(\omega) & 0 &  \beta_{yz}(\omega)\\
0 &  \beta_{zy}(\omega) & 0
\end{array}\right)
\end{eqnarray}

Then tensor matrix of $C_{2v}=C_2 \cap C^{\prime}_{1h}$ is the intersection of the tensor matrixs $C_2$ and $C^{\prime}_{1h}$, and we can get

\begin{eqnarray}
C_{2v}: \left(\begin{array}{ccc}
 0 & \beta_{xy}(\omega) & 0\\
 \beta_{yx}(\omega) & 0 & 0\\
 0 & 0 & 0
\end{array}\right)
\end{eqnarray}

Following this way, we can get the matrix of tensor $\beta_{i j}(\omega)$ for all nomagnetic point groups without inversion symmetry, shown in the Table.\ref{beta}, our results are consistent with the Ref.\cite{Sturman1992}. There are three special point groups $C_{3h}$, $D_{3h}$ and $T_d$, where all matrix terms of tensor $\beta_{i j}(\omega)$ are zero, and the reason is that the rotation and mirror symmetries together forbid tensor $\beta_{i j}(\omega)$. For examples, the point group $C_{3h}$ can be written as $C_3\otimes C_{1h}$, and by using tensor matrix of $C_3$ and $C_{1h}$, the tensor $\beta_{i j}(\omega)$ can be forbidden.

\renewcommand\arraystretch{2.0}
\begin{table*}
\caption{\label{beta} The classification of CPGE tensor $\beta_{i j}(\omega)$ for all nomagnetic point groups, the principal axis in crystal symmetries is along the z direction.}
\setlength{\tabcolsep}{1.5mm}{
\begin{tabular}{|c|c|c|}
  \hline
 Crystal system & Point group & $\boldsymbol{\beta}(\omega)$ tensor   \\\hline
Triclinic& $C_1$ &
$C_1$:$\left(\begin{array}{ccc}
 \beta_{xx}(\omega) & \beta_{xy}(\omega) & \beta_{xz}(\omega)\\
 \beta_{yx}(\omega) & \beta_{yy}(\omega) & \beta_{yz}(\omega)\\
 \beta_{zx}(\omega) & \beta_{zy}(\omega) & \beta_{zz}(\omega)
\end{array}\right)$\\\hline
Monoclinic& $C_2$,~ $C_{1h}$ & $C_2$:$\left(\begin{array}{ccc}
 \beta_{xx}(\omega) & \beta_{xy}(\omega) & 0\\
 \beta_{yx}(\omega) & \beta_{yy}(\omega) & 0\\
 0 & 0 & \beta_{zz}(\omega)
\end{array}\right)$,~$C_{1h}$:$\left(\begin{array}{ccc}
 0 & 0 & \beta_{xz}(\omega)\\
 0 & 0 & \beta_{yz}(\omega)\\
 \beta_{zx}(\omega) & \beta_{zy}(\omega) & 0
\end{array}\right)$  \\\hline
Orthorhombic& $D_2$,~$C_{2v}$  & $D_2$:$\left(\begin{array}{ccc}
 \beta_{xx}(\omega) & 0 & 0\\
 0 & \beta_{yy}(\omega) & 0\\
 0 & 0 & \beta_{zz}(\omega)
\end{array}\right)$,~$C_{2v}$:$\left(\begin{array}{ccc}
 0 & \beta_{xy}(\omega) & 0\\
 \beta_{yx}(\omega) & 0 & 0\\
 0 & 0 & 0
\end{array}\right)$  \\\hline
Tetragonal & $C_4$,~$S_{4}$,~$D_4$ & $C_4$:$\left(\begin{array}{ccc}
 \beta_{xx}(\omega) & \beta_{xy}(\omega) & 0\\
 -\beta_{xy}(\omega) & \beta_{xx}(\omega) & 0\\
 0 & 0 & \beta_{zz}(\omega)
\end{array}\right)$,~$S_{4}$:$\left(\begin{array}{ccc}
 \beta_{xx}(\omega) & \beta_{xy}(\omega) & 0\\
 \beta_{xy}(\omega) & -\beta_{xx}(\omega) & 0\\
 0 & 0 & 0
\end{array}\right)$,~$D_4$:$\left(\begin{array}{ccc}
 \beta_{xx}(\omega) & 0 & 0\\
 0 & \beta_{xx}(\omega) & 0\\
 0 & 0 & \beta_{zz}(\omega)
\end{array}\right)$     \\
       &$C_{4v}$,~$D_{2d}$  & $C_{4v}$:$\left(\begin{array}{ccc}
 0 & \beta_{xy}(\omega) & 0\\
 -\beta_{xy}(\omega) & 0 & 0\\
 0 & 0 & 0
\end{array}\right)$,~$D_{2d}$:$\left(\begin{array}{ccc}
 \beta_{xx}(\omega) & 0 & 0\\
 0 & -\beta_{xx}(\omega) & 0\\
 0 & 0 & 0
\end{array}\right)$     \\\hline
Trigonal & $C_3$,~$D_3$,~$C_{3v}$ &  $C_3$:$\left(\begin{array}{ccc}
 \beta_{xx}(\omega) & \beta_{xy}(\omega) & 0\\
 -\beta_{xy}(\omega) & \beta_{xx}(\omega) & 0\\
 0 & 0 & \beta_{zz}(\omega)
\end{array}\right)$,~$D_{3}$:$\left(\begin{array}{ccc}
 \beta_{xx}(\omega) & 0 & 0\\
 0 & \beta_{xx}(\omega) & 0\\
 0 & 0 & \beta_{zz}(\omega)
\end{array}\right)$,~$C_{3v}$:$\left(\begin{array}{ccc}
 0 & \beta_{xy}(\omega) & 0\\
 -\beta_{xy}(\omega) & 0 & 0\\
 0 & 0 & 0
\end{array}\right)$    \\\hline
Hexagonal   & $C_6$,~$C_{3h}$,~$D_6$ & $C_6$:$\left(\begin{array}{ccc}
 \beta_{xx}(\omega) & \beta_{xy}(\omega) & 0\\
 -\beta_{xy}(\omega) & \beta_{xx}(\omega) & 0\\
 0 & 0 & \beta_{zz}(\omega)
\end{array}\right)$,~$C_{3h}$:$\left(\begin{array}{ccc}
 0 & 0 & 0\\
 0 & 0 & 0\\
 0 & 0 & 0
\end{array}\right)$,~$D_{6}$:$\left(\begin{array}{ccc}
 \beta_{xx}(\omega) & 0 & 0\\
 0 & \beta_{xx}(\omega) & 0\\
 0 & 0 & \beta_{zz}(\omega)
\end{array}\right)$   \\
     & $C_{6v}$,~$D_{3h}$ & $C_{6v}$:$\left(\begin{array}{ccc}
 0 & \beta_{xy}(\omega) & 0\\
 -\beta_{xy}(\omega)& 0 & 0\\
 0 & 0 & 0
\end{array}\right)$,~$D_{3h}$:$\left(\begin{array}{ccc}
 0 & 0 & 0\\
 0 & 0 & 0\\
 0 & 0 & 0
\end{array}\right)$  \\\hline
Cubic& $T$,~$O$,~$T_{d}$ & $T$:$\left(\begin{array}{ccc}
 \beta_{xx}(\omega) & 0 & 0\\
 0 & \beta_{xx}(\omega) & 0\\
 0 & 0 & \beta_{xx}(\omega)
\end{array}\right)$,~$O$:$\left(\begin{array}{ccc}
 \beta_{xx}(\omega) & 0 & 0\\
 0 & \beta_{xx}(\omega) & 0\\
 0 & 0 & \beta_{xx}(\omega)
\end{array}\right)$,~$T_{d}$:$\left(\begin{array}{ccc}
 0 & 0 & 0\\
 0 & 0 & 0\\
 0 & 0 & 0
\end{array}\right)$    \\\hline
\end{tabular}}
\end{table*}

\begin{figure*}
\centerline{\includegraphics[width=0.7\textwidth]{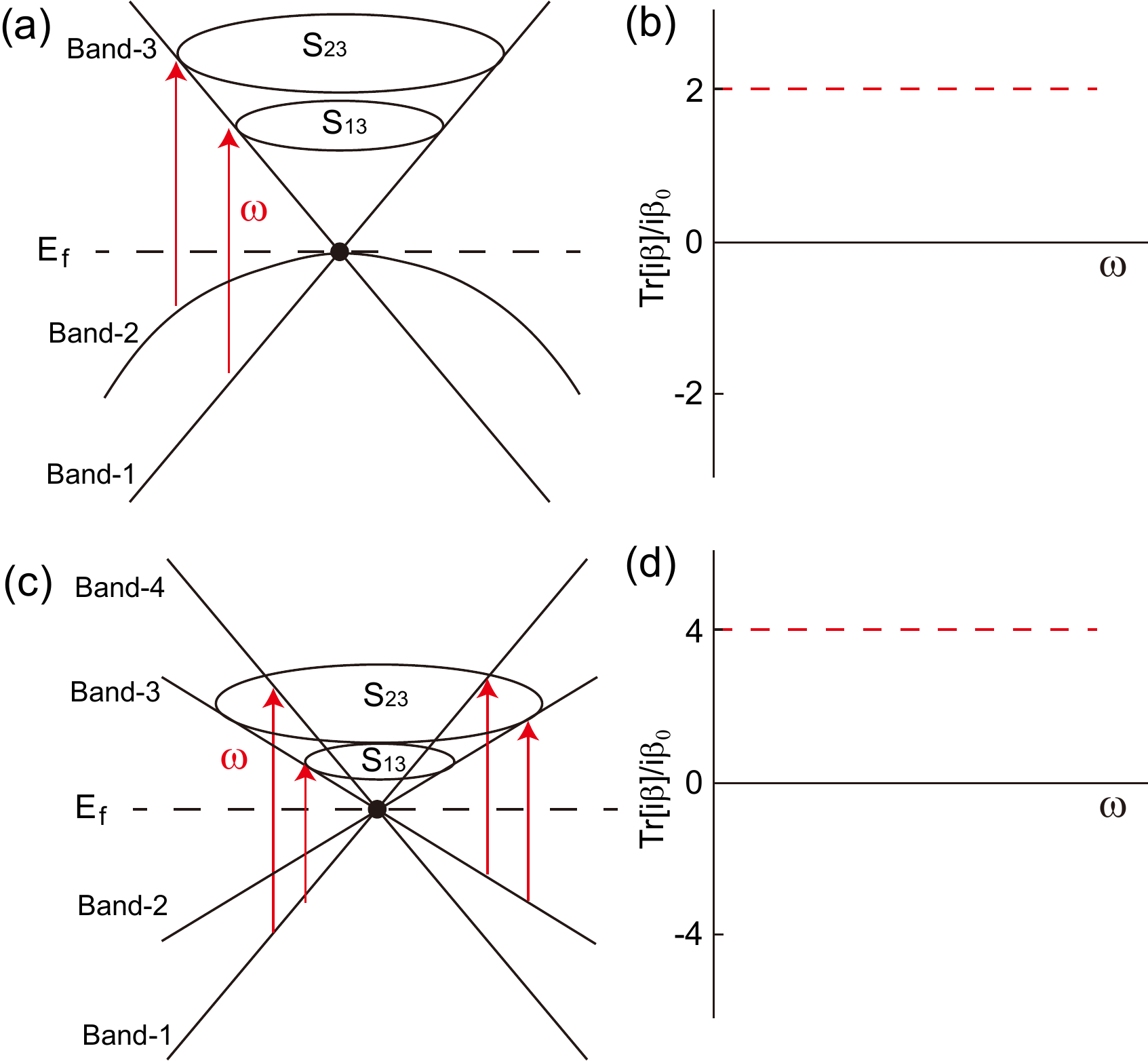}}
\caption{(color online) (a) and (b) Band structure of Type-I triply degenerate point in the three-band effective $k\cdot p$ model, and the corresponding trace of CPGE tensor as a function of frequency $\hbar\omega$. (c) and (d) Band structure of fourfold degenerate point  in the four-band effective $k\cdot p$ model, and the corresponding trace of CPGE tensor as a function of frequency $\hbar\omega$. The surface $\boldsymbol{S}_{12}$ and $\boldsymbol{S}_{23}$ are closed, and the chemical potential $E_f$ is indicated by by a dashed horizontal line.
\label{fig2} }
\end{figure*}

.

\section{Quantized CPGE in topological semimetals}\label{S2}

The noncentrosymmetric topological semimetals with nontrivial topological invariants, such as Weyl point and unconventional multifold fermions, always attract extensive attentions. Those multifold fermions carry monopoles of Berry curvature, and Chern number can be defined by the flux of the Berry curvature through a closed surface wrapping multifold fermions in momentum space, which lead to a series of exotic effects such as
large anomalous Hall, spin Hall effect\cite{Enk_2017,Wangqi2017,sun2016} and nonlinear optical responses\cite{Zhong2016,Ma2015,Chan2017,Ma2017,Sipe2000,Juan2017,Felix2018}. With such topological band structures, it is a natural question: Is there a quantized signal in topological semimetals? The Ref.\cite{Juan2017,Chang2017,Felix2018} give answer: the CPGE can become a quantized response in the noncentrosymmetric topological semimetals, which depend on the Chern number of multifold fermions\cite{Juan2017,Felix2018}. The following will give detailed derivation of the quantized CPGE trace.

Let's review Chern number in the topological semimetals, and the topological charge of multifold fermions can be characterized by the Chern number

\begin{eqnarray}
\mathcal{C}_{n}=\frac{1}{2 \pi} \oint_{S} \boldsymbol{\Omega}_{n}(\boldsymbol{k}) \cdot d \boldsymbol{S}_n
\end{eqnarray}

Where $\boldsymbol{S}_n$ is a closed surface of band $n$ enclosing
the multifold fermions, and $\boldsymbol{\Omega}_{n}(\boldsymbol{k})=\boldsymbol{\nabla}_{\boldsymbol{k}} \times\langle\psi_{n}(\boldsymbol{k})|i \boldsymbol{\nabla}_{\boldsymbol{k}}| \psi_{n}(\boldsymbol{k})\rangle$
is the Berry curvature of band $n$

In spherical coordinates, the CPGE trace becomes

\begin{eqnarray}
&&\beta(\omega)\nonumber
\\
&=&\frac{4 \pi^{2} \beta_{0}}{V}\sum_{i,\boldsymbol{k}, n, m}  \Delta_{\boldsymbol{k}, n m}^{i} \epsilon_{i k l}r_{\boldsymbol{k}, n m}^{k}  r_{\boldsymbol{k}, m n}^{l} \delta(\hbar \omega-E_{\boldsymbol{k}, m n})\nonumber
\\
&=&\frac{4 \pi^{2} \beta_{0}}{V}\sum_{\boldsymbol{k}, n, m}  \Delta_{\boldsymbol{k}, n m}^{i} R^{i}_{\boldsymbol{k},n m}\delta(\hbar \omega-E_{\boldsymbol{k}, m n})\nonumber
\\
&=&4 \pi^{2} \beta_{0} \sum_{\boldsymbol{k}, n, m} \int \frac{\mathrm{d} k dS}{(2 \pi)^{3}}  \partial_{\boldsymbol{k}} E_{\boldsymbol{k},n m}\cdot{R}_{\boldsymbol{k},n m} \delta(\hbar \omega-E_{\boldsymbol{k}, m n})\nonumber
\\
&=&4 \pi^{2} \beta_{0} \sum_{n, m}\int \frac{\mathrm{d} k dS}{(2 \pi)^{3}}  \partial_{\hat{k}} E_{\boldsymbol{k},n m}  R_{\hat{k},n m} \delta(\hbar \omega-E_{\boldsymbol{k}, m n})\nonumber
\\
&=&4 \pi^{2} \beta_{0} \sum_{n, m}\int \frac{ \mathrm{d} (E_{\boldsymbol{k},n m}) dS}{(2 \pi)^{3}}   R_{\hat{k},n m} \delta(\hbar \omega-E_{\boldsymbol{k}, m n})\nonumber
\\
&=&\frac{\beta_0}{2\pi}  \sum_{n, m} \int d \boldsymbol{S}_{n m} \cdot \boldsymbol{R}_{n m}
\end{eqnarray}

Here, in spherical coordinates $\partial_{\boldsymbol{k}} E_{\boldsymbol{k},n m}=\partial_{k} E_{\boldsymbol{k},n m}\hat{k}+\frac{1}{k} \partial_{\theta} E_{\boldsymbol{k},n m} \hat{\theta}+\frac{1}{k \sin \theta} \partial_{\phi} E_{\boldsymbol{k},n m} \hat{\phi}$ and ${R}_{\boldsymbol{k},n m}=R_{\hat{k},n m}\hat{k}+R_{\theta,n m}\hat{\theta}+R_{\phi,n m}\hat{\phi}$. We set $f^{\boldsymbol{k}}_{nm}=1$ and $\delta$ function is trivially integrated. ${R}_{\boldsymbol{k},n m} $has only the radial component in spherical coordinates for multifold degenerate points with linear dispersion\cite{Felix2018}. The band indexes $n$ and $m$ represent all occupied and unoccupied states, respectively. $d\boldsymbol{S}$ denotes the oriented surface element normal to $\boldsymbol{S}$. The relation between $\boldsymbol{R}_{\mathbf{k},n m}$ and Berry curvature is $\boldsymbol{\Omega}_{n}(\boldsymbol{k})=i\sum_{m\neq n}\boldsymbol{R}_{\mathbf{k},n m}$, and hence the CPGE trace can be physically understood as the Berry flux penetrating through $\boldsymbol{S}$. We notice that there is a different coefficient with the Ref.\cite{Felix2018}. In the following, Based on effective $k\cdot p$ models of multifold fermions, the quantized CPGE trace will be discussed in all kinds of noncentrosymmetric topological semimetals.

For a single type-I Weyl point with Chern number $C=\pm1$, the Fermi level is located at the Weyl point, shown in Fig.\ref{fig1}(a). In the two-band effective $k\cdot p$ model, the CPGE trace $\beta(\omega)$ becomes

\begin{eqnarray}
\beta(\omega)&=&\frac{\beta_0}{2\pi} \int d \boldsymbol{S}_{12} \cdot \boldsymbol{R}_{12}\nonumber
\\
&=&-i \frac{\beta_0}{2\pi} \int d \boldsymbol{S}_{12} \cdot \boldsymbol{\Omega}_{1}
\nonumber
\\
&=&-i C_{1} \beta_{0}=\pm i\beta_{0},
\end{eqnarray}

where $\boldsymbol{R}_{12}$ is equal to Berry curvature $\boldsymbol{\Omega}_{1}$ in the two-band model according to $\boldsymbol{\Omega}_{n}(\boldsymbol{k})=i\sum_{m\neq n}\boldsymbol{R}_{\mathbf{k},n m}$, and $C_1$ is Chern number of the occupied band. Since the surface $\boldsymbol{S}_{12}$ is closed and wraps Weyl point, we can obtain the third line. Hence, the CPGE trace is quantized for a type-I Weyl point, shown in Fig.\ref{fig1}(b). However, for a type-II Weyl point shown in Fig.\ref{fig1}(c), the surface $\boldsymbol{S}_{12}$ cannot be closed due to the overtilted band crossing, and  the CPGE trace is never quantized.

For the triply degenerate points(TDPs), they can be classified as type-I,type-II and type-III, and the topological charges are $\pm2$, $\pm1$ and 0, respectively\cite{Hu2018}. Simiar to type-II Weyl points, the type-II and type-III TDPs also overtilted, which can lead to open surface $\boldsymbol{S}$, indicating that  the CPGE trace is never quantized. Then, we only discuss type-I TDP, where the Chern numbers of band-1(band-3) are $\pm2$ and the band-2 is zero, shown in Fig.\ref{fig2}(a). If the Fermi level is located at the type-I TDP, in the three-band effective $k\cdot p$ model, the CPGE trace $\beta(\omega)$ becomes

\begin{eqnarray}
\beta(\omega)&=& \frac{\beta_0}{2\pi}(\int d \boldsymbol{S}_{13} \cdot \boldsymbol{R}_{13}+\int d \boldsymbol{S}_{23} \cdot \boldsymbol{R}_{23}),\nonumber \\
&=& \frac{\beta_0}{2\pi}(-i \int d \boldsymbol{S}_{13} \cdot \boldsymbol{\Omega}_{1}-i \int d \boldsymbol{S}_{23} \cdot \boldsymbol{\Omega}_{2},\nonumber \\
&-&\int d \boldsymbol{S}_{13} \cdot\boldsymbol{R}_{12}-\int d \boldsymbol{S}_{23} \cdot\boldsymbol{R}_{21}),\nonumber \\
&=& -i \frac{\beta_0}{2\pi}(\int d \boldsymbol{S}_{13} \cdot \boldsymbol{\Omega}_{1}+\int d \boldsymbol{S}_{23} \cdot \boldsymbol{\Omega}_{2}),\nonumber \\
&=& -i \beta_{0} (C_{1}+ C_{2})=\pm 2i\beta_{0},
\end{eqnarray}

where $C_1=\pm 2$ and $C_2=0$ are Chern number of the occupied band-1 and band-2, and the surface $\boldsymbol{S}_{12}$ and $\boldsymbol{S}_{23}$  are closed. Hence, the CPGE trace is quantized for a type-I TDP, shown in Fig.\ref{fig2}(b), where we take $C_1=-2$.

For a Rarita-Schwinger-Weyl(RSW) fermion\cite{Bradlyn2016,Rarita1941,Liang2016,Ezawa2016}, a spin RSW fermion acts as a monopole with topological charge $\pm4$, shown in Fig.\ref{fig2}(c). We can label bands from top to bottom as band-$n$ ($n$= 1, 2, 3, 4), and the corresponding Chern numbers are $C_{1/4}=\pm3$ and $C_{2/3}=\pm1$.

\begin{eqnarray}
\beta(\omega)&=& \frac{\beta_0}{2\pi}(\int d \boldsymbol{S}_{13} \cdot \boldsymbol{R}_{13}+\int d \boldsymbol{S}_{14} \cdot \boldsymbol{R}_{14},\nonumber \\
&+&\int d \boldsymbol{S}_{23} \cdot \boldsymbol{R}_{23}+\int d \boldsymbol{S}_{24} \cdot \boldsymbol{R}_{24}),\nonumber \\
&=& \frac{\beta_0}{2\pi}(-i \int d \boldsymbol{S}_{13} \cdot \boldsymbol{\Omega}_{1}-i \int d \boldsymbol{S}_{23} \cdot \boldsymbol{\Omega}_{2},\nonumber \\
&-&\int d \boldsymbol{S}_{13} \cdot(\boldsymbol{R}_{12}+\boldsymbol{R}_{14})+\int d \boldsymbol{S}_{14} \cdot \boldsymbol{R}_{14},\nonumber \\
&-&\int d \boldsymbol{S}_{23} \cdot\left(\boldsymbol{R}_{21}+\boldsymbol{R}_{24}\right)+\int d \boldsymbol{S}_{24} \cdot \boldsymbol{R}_{24}),\nonumber \\
&=& -i\frac{\beta_0}{2\pi}(\int d \boldsymbol{S}_{13} \cdot \boldsymbol{\Omega}_{1}+\int d \boldsymbol{S}_{23} \cdot \boldsymbol{\Omega}_{2}),\nonumber \\
&=& -i \beta_{0}(C_{1}+C_{2})=\pm 4i\beta_{0},
\end{eqnarray}

where $C_1=\pm 3$ and $C_2=\pm1$ are Chern number of the occupied band-1 and band-2, and the surface $\boldsymbol{S}_{12}$ and $\boldsymbol{S}_{23}$  are closed. Hence, the CPGE trace is quantized for a RSW fermion, shown in Fig.\ref{fig2}(d), where we take $C_1=-3$ and $C_2=-1$.

\begin{figure*}
\centerline{\includegraphics[width=0.9\textwidth]{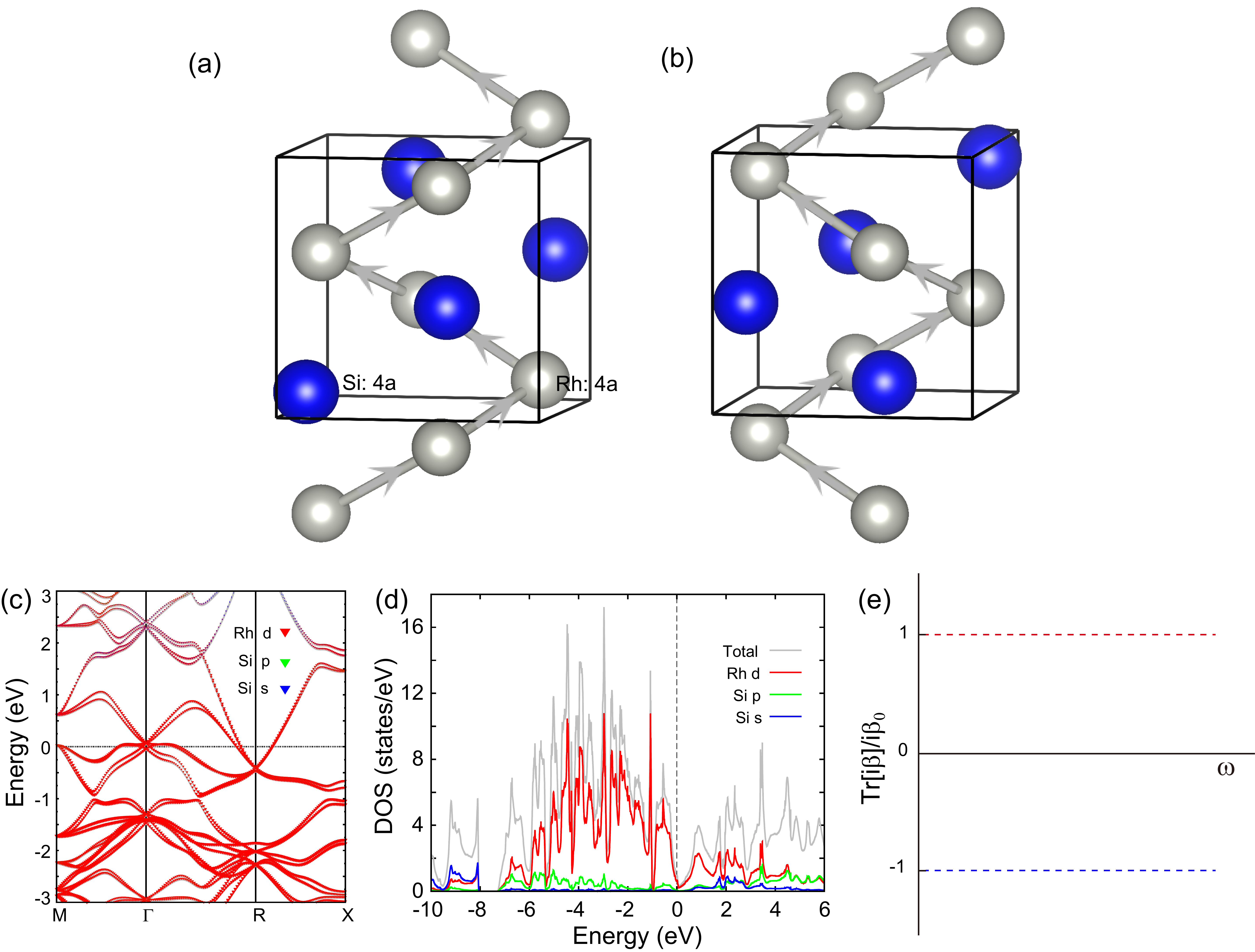}}
\caption{(color online) (a) and (b) Two chiral crystal structure of RhSi connected by inversion symmetry. (c) and (d) Band structures and DOS for RhSi with SOC. (e) Schematic of CPGE trace in two chiral crystal structures, if one of two chiral crystal structures is +1 and the other one is -1. The orbital characters of bands are represented by different colors.}
\label{RhSi_1}
\end{figure*}

\section{chiral topological semimetal RhSi materials family}\label{S3}

In this section, we will review all chiral topological semimetals in RhSi family, including CoSi, PdGa, PtGa, PtAl and RhSn, which may be the promising candidates to exhibit a quantized CPGE. For completeness, we analyze RhSi family in terms of crystal structure, crystal symmetries, band structure and CPGE tensor, where we use RhSi as an example to introduce crystal and band structures.

The crystal structure of RhSi family with FeSi-type structure is shown in Fig.\ref{RhSi_1}(a) and (b), where Rh and Si atoms occupy the same Wyckoff position 4a $\{(x, x, x), (-x+1/2, -x, x+1/2), (-x, x+1/2, -x+1/2), (x+1/2, -x+1/2, -x)\}$. The space group G=P2$_1$3 (No.198) of RhSi family is nonsymmorphic with point group $T$. A space group is symmorphic if the translation part in the Seitz operators\cite{Seitz1935} $\{\alpha|{\bf \tau}\}$ are all crystal translations, where $\alpha$ is a point group operation and ${\bf \tau}$ is a spatial translation. However, it is nonsymmorphic if there is at least one Seitz operator whose translation part is non-crystal translation and the noncrystal translation can not be eliminated by changing the origin point to a proper position\cite{Liao2012}. For examples, the symmetry operation $\{\alpha \mid v(\alpha)\}$ is the nonsymmorphic opertations,where $v(\alpha)$ is a fractional translation. When the origin point is translated by a vector $\vec{b}$, the non-crystal translation operators would change to $\{E \mid \vec{b}\}$$\{\alpha \mid v(\alpha)\}$$\{E \mid \vec{b}\}^{-1}$=$\{\alpha \mid v(\alpha)+\vec{b}-{\alpha}\vec{b}\}$. If there is a proper origin $\vec{b}$ that can make $v(\alpha)+\vec{b}-\alpha\vec{b}$ a crystal translation, the operator is symmorphic; Otherwise, it is nonsymmorphic. In the RhSi family, the nonsymmorphic symmetry operations are $\tilde{C}_{2x}=\{C_{2x}|1/2, 1/2, 0\}$, $\tilde{C}_{2y}=\{C_{2y}|0, 1/2, 1/2\}$, and $\tilde{C}_{2z}=\{C_{2y}|1/2, 0, 1/2\}$. The nonsymmorphic operators can lead to high degeneracy on the Brillourin zone boundary, which can protect unconventional quasi-particles.

\begin{figure*}
\centerline{\includegraphics[width=0.9\textwidth]{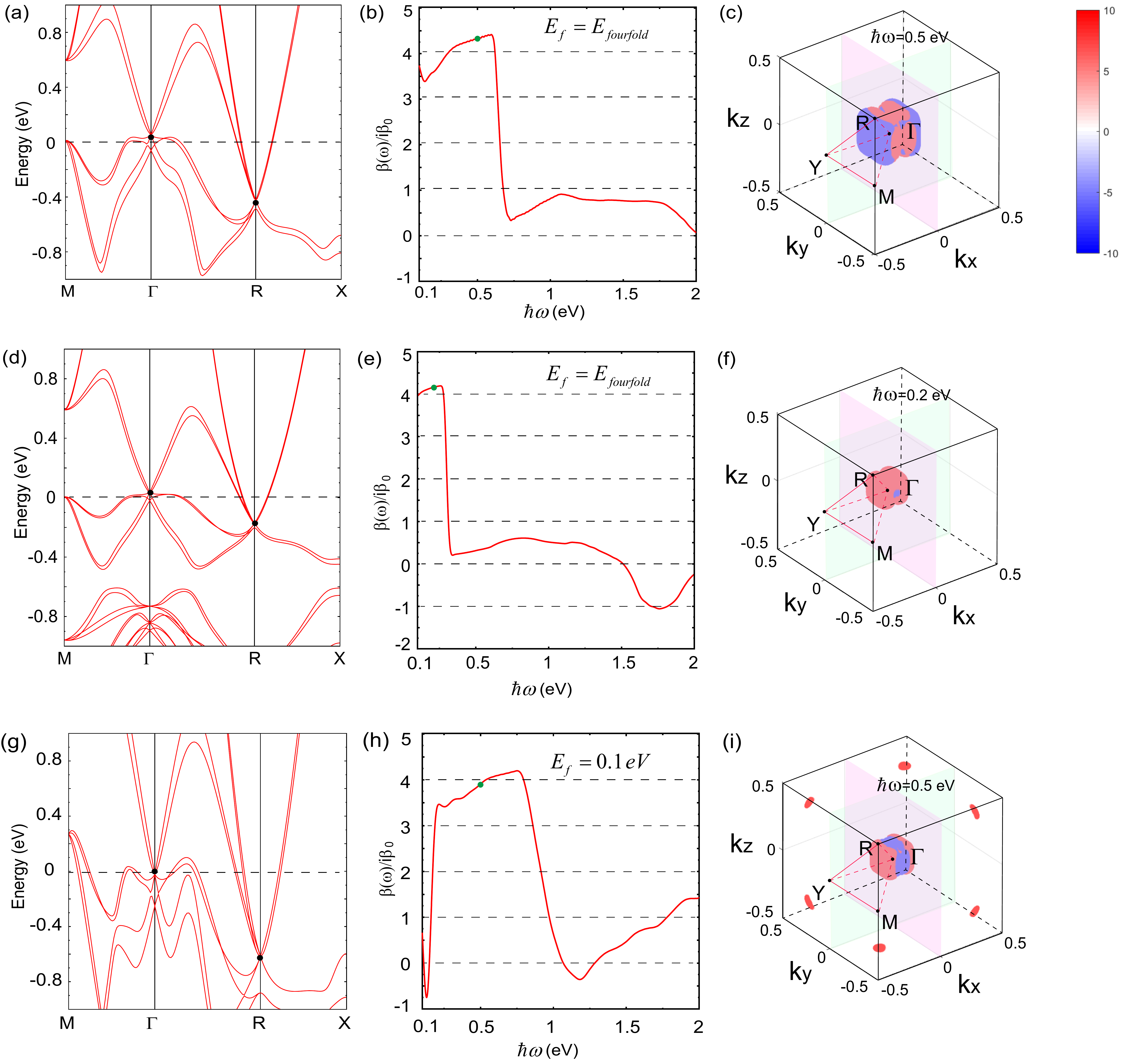}}
\caption{(color online) Band structures with SOC in the paramagnetic state, the trace of
the CPGE tensor as a function of frequency, and momentum distribution of the CPGE tensor
from all bands for (a)-(c) RhSi at frequency $\hbar\omega=0.5eV$, (d)-(f) PtAl at frequency $\hbar\omega=0.2eV$, and (g)-(i) CoSi at frequency $\hbar\omega=0.5eV$. The green dots show corresponding the size of CPGE traces. In addition to the nontrivial bands that form fourfold fermion, the momentum distribution also comes from other trivial bands, and there are red and blue in the momentum distribution.
\label{RhSi_2} }
\end{figure*}

The band structure and density of states (DOS) of RhSi with SOC are displayed in Fig.\ref{RhSi_1}(c) and (d). The Si $p$- and $s$-orbitals are far away from Fermi level, and near Fermi level the valence and conduction bands are predominantly attributed to Rh-$d$ orbitals. The most prominent feature is two types of symmetry-protected band crossings with linear dispersion, one at $\Gamma$ point with sixfold degeneracy and the other at R point with eightfold degeneracy. When SOC is included, the sixfold degeneracy split into a fourfold degeneracy and a twofold degeneracy, while the eightfold degeneracy can lead to a sixfold degeneracy and a twofold degeneracy, shown in Fig.\ref{RhSi_2}(c). The fourfold degeneracy is protected by point group T and time-reversal symmetry, while the sixfold degeneracy is protected by non-symmorphic symmetries and time-reversal symmetry\cite{Bradlyn2016}. Because fourfold and sixfold degenerate points
are not symmetrically related, the energies of these points can be different. The fourfold
degeneracy at $\Gamma$ point is described by a spin-3/2 fermion with Chern number C=3, 1, -1, -3
for the four bands\cite{Chang2017,Tang2017}, and the sixfold degeneracy at $R$ point is double
spin-1 fermion with C=2, 2, 0, 0, -2,-2 for the six bands, which satisfies Nielsen-Ninomiya
theorem\cite{Nielsen1,Nielsen2,Nielsen3}.

\begin{figure*}
\centerline{\includegraphics[width=0.9\textwidth]{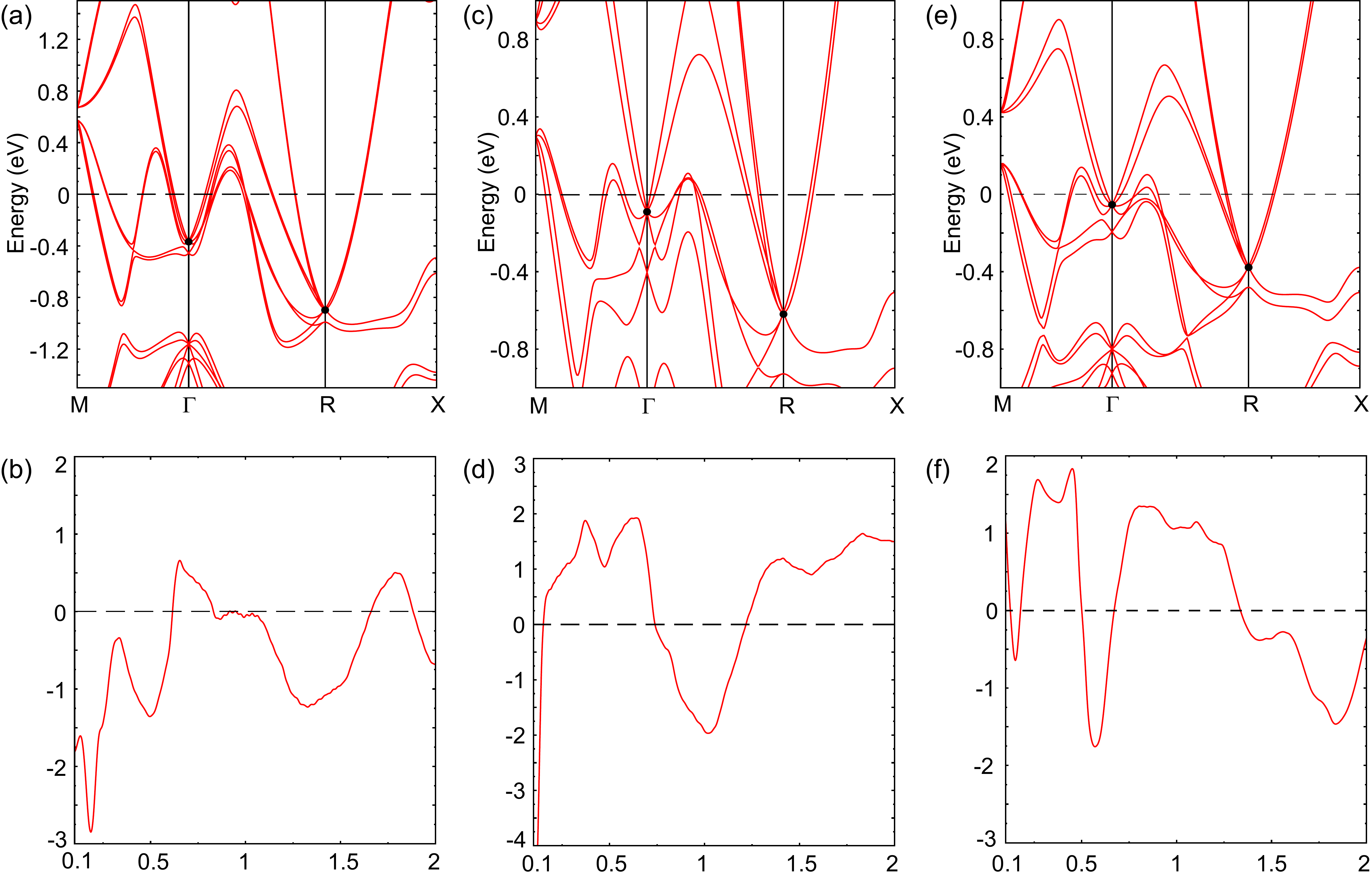}}
\caption{(color online) Band structures with SOC in the paramagnetic state and the trace of
the CPGE tensor as a function of frequency for (a)-(b) PdGa ,(c)-(d) PtGa and (e)-(f) RhSn.
 \label{RhSi_3}}
\end{figure*}

Since the quantized CPGE trace is based on the effective $k\cdot p$ models of multifold fermions with linear dispersion, several factors can modify the quantization in the real materials. Before we give detailed results about quantized CPGE in RhSi family, the difference between real material and effective model will be discussed\cite{Le2020}.  In order to have quantized CPGE, the multifold fermions with opposites topological charges are not the same energy. Hence, if one fermion is located at Fermi level, the other one away from Fermi level can contribute to fermi surface. Such as, in the Fig.\ref{RhSi_1}(c), when Fermi level is at fourfold fermion at $\Gamma$ point, and the sixfold fermion at $R$ point can contribute to electron pockets. If we consider quantized CPGE trace from fourfold fermion, the electron pockets near $R$ point can modify it. When mutifold fermion do not have perfect linear dispersion, the quantization maybe absent. Fig.\ref{RhSi_3} show the band structures of PdGa, PtGa and RhSn, where the fourfold fermions at $\Gamma$ point do not have perfect linear dispersion, indicating that these materials can not host quantized CPGE.

Depending on whether there is the quantization of CPGE, the RhSi family can be separated into two groups. The first group is RhSi, PtAl and CoSi with nearly quantized CPGE trace, while the second group includes PdGa, PtGa and RhSn without quantization. Fig.\ref{RhSi_2} show band structures with SOC, the trace of CPGE tensor as a function of frequency, and momentum distribution of the CPGE tensor for the first group materials, and the following will give detailed analysis.

For RhSi, we focus on Fermi level at fourfold degeneracy, and a nearly plateau close to 4$\beta_0$ with the frequency from ~0.1 to ~0.6 eV is shown in Fig.\ref{RhSi_2}(b). The optical transitions from the sixfold fermion are forbidden due to Pauli blocking in the frequency range from ~0.1 to ~0.6 eV, and the nontrivial bands that form fourfold fermion play a major role in CPGE. When optical frequency is above 0.6 eV, the nontrivial bands from sixfold fermion can begin to contribute CPGE, and cancel the contribution from fourfold fermion, indicating that the CPGE trace is close to zero from ~0.7 to ~2 eV. Since contributions to CPGE from other trivial bands cannot be ruled out, and the plateau close to 4$\beta_0$ range from ~0.1 to ~0.6 eV and exact zero range from 0.6 to 2 eV are absent in the DFT calculations. Then, we calculate the momentum distribution of CPGE from all bands in the BZ at frequency $\hbar\omega=0.5 eV$ to further analyze CPGE, shown in Fig.\ref{RhSi_2}(c). One can see the CPGE completely contribute to electric optical transitions of bands near $\Gamma$ point. In addition to the nontrivial bands that form fourfold fermion, the momentum distribution also comes from other trivial bands, and there are red and blue in the momentum
distribution. Hence, a nearly plateau with the frequency from ~0.1 to ~0.6 eV is contributed by fourfold fermion.

Fig.\ref{RhSi_2}(d) shows band structures of CoSi with SOC, and the fourfold degeneracy at $\Gamma$ point
is near Fermi level, while the sixfold degeneracy at $R$ point is far away. Similar to RhSi, we also focus on Fermi level at fourfold fermion, and a nearly plateau close to 4$\beta_0$ is shown in Fig.\ref{RhSi_2}(e). However, compared with RhSi, the frequency range of plateau is very narrow, which is from ~0.1 to ~0.2 eV, and the reason is that the energy difference between fourfold and sixfold degeneracy is small. When  the frequency is less than 0.2 eV, the optical transitions from the sixfold fermion are forbidden due to Pauli blocking and the fourfold fermion mainly contribute to CPGE. When optical frequency $\omega$ is more than 0.2 eV, the sixfold fermion begins to contribute CPGE, and cancel the contribution from fourfold fermion, indicating that the CPGE trace is close to zero from ~0.7 to ~2 eV. Then, we calculate the momentum distribution of CPGE trace from all bands in the BZ at frequency $\hbar\omega=0.2 eV$, shown in Fig.\ref{RhSi_2}(f), indicating that the CPGE completely contribute to electric optical transitions of bands near $\Gamma$ point. Hence, a nearly plateau with the frequency from ~0.1 to ~0.2 eV is contributed by fourfold fermion at $\Gamma$ point.

Fig.\ref{RhSi_2}(g) shows band structures with SOC of PtAl in the paramagnetic states, where the fourfold and sixfold fermions denoted by black points are close to and away from the Fermi level, respectively. Compared with RhSi and CoSi, the band dispersion of fourfold degeneracy is not perfect linear, and then we set the Fermi level at 0.1eV.  A large quantized platform can be obtain, shown in Fig.\ref{RhSi_2}(h). Then, we calculate the momentum distribution of CPGE from all bands in the BZ at frequency $\hbar\omega=0.5 eV$ to further analyze CPGE trace, shown in Fig.\ref{RhSi_2}(i), indicating that the CPGE completely contribute to electric optical transitions of bands near $\Gamma$ point. Hence, a nearly plateau with the frequency from ~0.1 to ~0.6 eV is contributed by fourfold fermion at $\Gamma$ point.

Fig.\ref{RhSi_3} shows band structures with SOC and the trace of CPGE tensor as a function of frequency for the second group materials, including PdGa, PtGa and RhSn. The common feature in the second group materials is that the dispersion of fourfold fermion at $\Gamma$ point is different from dispersion in the $k\cdot p$ model shown in Fig.\ref{fig2}(c). Fig.\ref{RhSi_3} shows the trace of CPGE as a function of frequency with Fermi level lying at the charge neutral point, and the quantized CPGE trace is absent.

Due to the lack of inversion and mirror symmetries in the chiral space groups, we can define chirality for materials with chiral space groups. A chiral crystal cannot be superposed on its mirror or inversion image, and the symmetry group of a chiral crystal contains only rotations, translations, and screw rotations. Fig.\ref{RhSi_1}(a) and (b) show two chiral crystal structure of RhSi connected by inversion. Since two chiral crystals have the same band structures, it is a natural question that how to detect chirality of topological semimetal materials\cite{Sun2020} in experiment. Firstly, the topological charge of multifold fermions is opposite in the two chiral semimetal materials. Secondly, the sign of CPGE can be determined by topological charge of multifold fermions according to Section.~\ref{S2}. Fig.\ref{RhSi_1}(e) shows schematic of CPGE trace in two chiral crystal structures with topological charge $\pm1$, when one of two chiral crystal structures is +1 and the other one is -1. Finally, the chirality of topological semimetal materials can be detected experimentally by measuring CPGE.

\section{Conclusion}\label{S4}

In summary, we review circular photogalvanic effect in the chiral multifold semimetals. Fistly, a table classifying matrix of CPGE tensors can be obtained by symmetric analysis. Secondly, based on $k \cdot p$ effective model with linear dispersion, the CPGE becomes a quantized response and depends on the Chern number of multifold fermions. Finally, according to ab-initio analysis for the quantized CPGE based on noninteractiong electronic structure, we analyze chiral topological semimetals in RhSi family, and only the first group can be the promising candidates to exhibit a quantized CPGE trace.

\end{document}